\documentclass{article}

\usepackage{arxiv}

\usepackage[utf8]{inputenc} 
\usepackage[T1]{fontenc}    
\usepackage{hyperref}       
\usepackage{url}            
\usepackage{booktabs}       
\usepackage{amsfonts}       
\usepackage{nicefrac}       
\usepackage{microtype}      
\usepackage{lipsum}

\usepackage{booktabs} 
\usepackage{subcaption}
\usepackage[font=bf]{caption}
\usepackage[utf8]{inputenc}
\usepackage[T1]{fontenc}   
\usepackage{dblfloatfix}    
\usepackage{graphicx}

\title{Emotional Avatars: The Interplay between Affect and Ownership of a Virtual Body}

\author{
  Aske Mottelson \\
  Department of Psychology\\
  University of Copenhagen\\
  Copenhagen, Denmark \\
  \texttt{amot@psy.ku.dk} \\
   \And
  Kasper Hornbæk \\
  Department of Computer Science\\
  University of Copenhagen\\
  Copenhagen, Denmark \\
  \texttt{kash@di.ku.dk} \\
}

\begin{document}
\maketitle

\begin{abstract}
Human bodies influence the owners' affect through posture, facial expressions, and movement. It remains unclear whether similar links between virtual bodies and affect exist. Such links could present design opportunities for virtual environments and advance our understanding of fundamental concepts of embodied VR.

An initial outside-the-lab between-subjects study using commodity equipment presented 207 participants with seven avatar manipulations, related to posture, facial expression, and speed. We conducted a lab-based between-subjects study using high-end VR equipment with 41 subjects to clarify affect's impact on body ownership.

The results show that some avatar manipulations can subtly influence affect. Study I found that facial manipulations emerged as most effective in this regard, particularly for positive affect. Also, body ownership showed a moderating influence on affect: in Study I body ownership varied with valence but not with arousal, and Study II showed body ownership to vary with positive but not with negative affect.
\end{abstract}


\section{Introduction}







The theory of embodied cognition 
suggests that \emph{cognitive processes are deeply rooted in the body's
interactions with the world}~\cite{wilson2002}, because of a
relation between bodily expression
of emotion and the way in which
emotional information is attended to~\cite{niedenthal2007, niedenthal2005}.
%
%
Here, we investigate the extent to which affective embodiment can be
observed with illusory ownership of a virtual body. Will a smiling virtual self cause joy and a frowning one prompt sadness?
We examine the causal
relationship between virtually inducing embodied affect and experiencing the relevant affective states.

Affect has been shown to be of great importance for many aspects of day-to-day life, among them cognitive performance, general health and well-being, creativity, decision-making processes, and social relationships~\cite{oxford}. Also, studies have shown that affect may influence perhaps the most fundamental VR concept: \emph{presence}, the \emph{sense of being there}~\cite{presenceemotion}.

\emph{Body ownership}, another key concept in VR, refers to the degree to which a virtual body is experienced as one's own body. Since bodies are demonstrably connected to affect, it seems worthwhile to investigate the relationship between virtual bodies and affect, and the link between affect and body ownership. However, no previous studies have explored these relationships. 

Through two user studies, this paper investigates the degree to which virtual bodies can modify affective responses, in addition to uncovering the role of affect for ownership of a virtual body. In summary, we present these findings as contributions:
\begin{itemize}
\item body-ownership illusions can influence affect (found in a large-scale user study with VR),
\item manipulation of facial features was most effective in influencing affect,
\item body ownership is a key component for positive affect,
\item body ownership showed to vary with the valence component of SAM, and
\item higher positive PANAS responses significantly increased the probability of high body ownership.
\end{itemize}

\section{Overview}

\subsection{What Affect Is}

The concepts of emotion, mood, and affect are sometimes mistakenly used interchangeably.  
We proceed from a view of affect inspired by Ekkekakis's efforts to untangle the conceptualization and measurement of affect~\cite{ekkekakis2013measurement}. He argued for distinguishing among three forms of affect: core affect, mood, and emotion. The most fundamental of these is core affect, underpinning moods and emotions. This is an evaluative feeling always available to the consciousness. Pleasure offers a clear example. Emotion, in contrast, depends on appraisal, and involves an object toward which the emotion is directed. Anger is one example. Finally, moods are long-term affective states in comparison to emotions; they are also less intense. Irritation is an apt example of a mood. In addition to presenting these basic categories, Ekkekakis argued that specific metrics employed in studies must operate from a particular view of affect (i.e., the entity composed of core affect, mood, and emotion). In particular, it must be clear whether the researcher is interested in dimensional or in categorical measures of affect. Consequently, we later explain the views of affect for both studies presented in the paper.

\begin{figure}[h!]
{ 
\centering
\frame{\includegraphics[width=\textwidth]{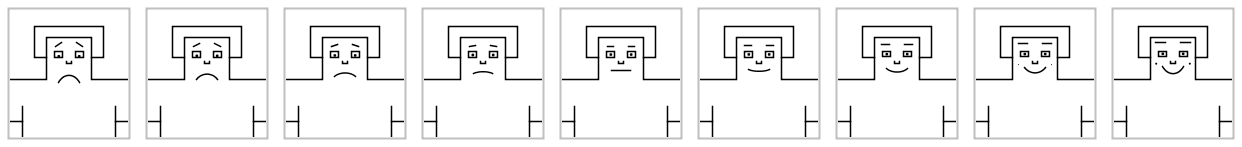}}\vspace{1.5mm}
\frame{\includegraphics[width=\textwidth]{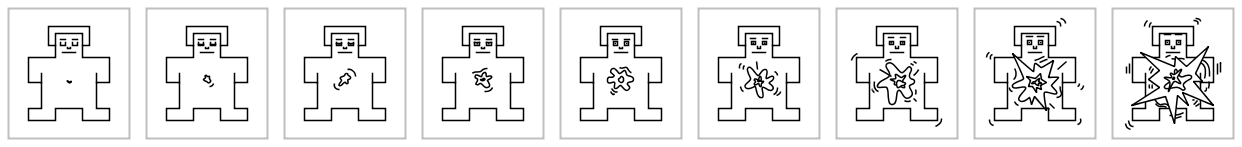}}
\caption{The nine-point Self-Assessment Manikin, measuring valence (unpleasant--pleasant, at the top) and arousal (deactivation--activation, at bottom).}
\label{fig:SAM}
}
\end{figure}

Both physiological and subjective measurements are commonly employed for estimation of affective responses. Especially popular among the latter are the Positive and Negative Affect Schedule (PANAS)~\cite{panas} and the Self-Assessment Manikin (SAM)~\cite{SAM}. The former considers positive and negative affect on independent scales, producing two summary scores from participants' rating of their emotional fit on a five-point scale with 10 positive and 10 negative words, while the SAM typically presents two scales, valence (pleasantness) and arousal (activation), using a pictorial manikin for representing the scores. There are versions with a five-, seven-, and nine-point scale (Figure~\ref{fig:SAM} shows a nine-point SAM).

\subsubsection{Manipulating Affect}

The process of altering affect experimentally is known as affect manipulation. It is sometimes colloquially referred to as mood induction or emotion elicitation (although this informal use is undesirable).  
Affect manipulation allows researchers to enforce specific affective responses from a sample than would have arisen if subjects had merely reported on their current affective state. 
Hence, this manipulation is an important psychological tool for understanding how mood, emotion, and core affect modify human cognition and behavior within the constraints of experimental scrutiny. 

There are several ways of manipulating affect (Westermann et al. give an overview~\cite{westermann1996}). Among the most commonly employed are film, IAPS, and Velten. These respectively involve showing affective movie clips, imagery, or asking participants to immerse themselves in an emotional story.

\subsection{Virtual Reality As an Affective Medium}

Research covering virtual reality and affect is surprisingly scarce,
notwithstanding the general consensus on affect's highly important influence on cognition and behavior, and
the recent proliferation of behavioral research using virtual reality. That said, some scholars have attempted to manipulate
affect via virtual environments~\cite{banos2006, felnhofer2015, toet2009, riva2007, Jun:2018}. We review that work below.

Ba\~{n}os et al.~\cite{banos2006} created a virtual environment capable of eliciting four discrete emotions. The environments depicted alterations of a city park that incorporated a multitude of
classical emotion-elicitation procedures, such as Velten, IAPS, and movie clips. The
full procedure, lasting 30 minutes, showed effective for discrete emotion elicitation.


Another emotion-eliciting virtual park was created by Felnhofer et
al.~\cite{felnhofer2015}. Five park scenarios were designed, each
eliciting a specific emotion: joy, anger, boredom, anxiety, or sadness.
The scenarios differed in lighting, coloring, and sound, along with their plant and animal types.
The authors concluded that virtual environments can be used to induce emotional states.

Riva et al.~\cite{riva2007} constructed their own park scenarios to elicit emotions
in VR. Their three parks shared the same structure
but differed in their aural and visual experience. These researchers too confirmed
the efficacy of VR as an affective medium.

To our knowledge, only Jun et al.~\cite{Jun:2018} have attempted to alter affect via avatar manipulations. They found that facial expressions of a virtual avatar can
modulate emotions and that greater presence is associated with higher valence.

\subsection{Body-Ownership Illusions}

Body-ownership illusions are a class of experiment in which the participants are led to believe that they are owners of another body, or part thereof~\cite{Petkova:2008}. These illusions are persuasive when the replacement bodies are virtual avatars made possible by VR~\cite{kilteni2015}. An optimal body-ownership illusion induces a high level of ownership of the virtual body, usually by means of visuo-motor synchrony between the physical and virtual body, sometimes alongside tactile feedback~\cite{rod}. 

While there is ongoing scholarly debate on how far avatar manipulations can be extended without distortions to body ownership, evidence thus far suggests that ownership can be maintained even with substantial alterations to the avatar, such as scaling the arms to three times the normal length~\cite{longarm}, rotating the body 15 degrees~\cite{bodydegrees}, or using a child's body for the avatar~\cite{Banakou:2013}.

A large body of research in this area is focused on manipulating avatars to study whether humans' perception of the world can be influenced through body alteration/replacement. For instance, owning a child avatar influences estimates of objects' size and expedites association of the self with childlike attributes~\cite{Banakou:2013}; similarly, owning a different-skin-toned avatar affects implicit racial bias~\cite{peck2013}.

The aforementioned body of work confirms the feasibility of changing body-oriented perceptions when participants are manipulated to believe the virtual body they see in VR is, in fact, their real body.





\subsection{Limitations of Previous Work}

While previous studies attest that VR scenarios can alter affect, this has been confined largely to park scenarios. The conditions involved are time-consuming, explicit in their purpose, and geared for a specific narrative. Most importantly, they also neglect embodiment as a part of the immersion.
This renders it hard to evaluate the processes leading to affect manipulation in embodied VR precisely.
While some evidence
suggests that one can influence affect and presence by manipulating the avatar's facial features~\cite{Jun:2018}, the connection between affect 
and the illusion of owning a virtual body remains unclear, as does whether other avatar manipulations may inform affect.
Some authors have shed light on affect's role for presence~\cite{banos2004}: the literature suggests a positive correlation between presence and affect.

We provide evidence that affect can be influenced by only bodies truly
thought to be ours. Additionally, we present a first study showing interplay between affect and body ownership. Our findings confirm the existence of a link between (positive) affect and virtual bodies.

\begin{figure*}[h!]
{ 
\centering
\begin{subfigure}{.195\textwidth}
  \centering
  \frame{\includegraphics[width=\textwidth]{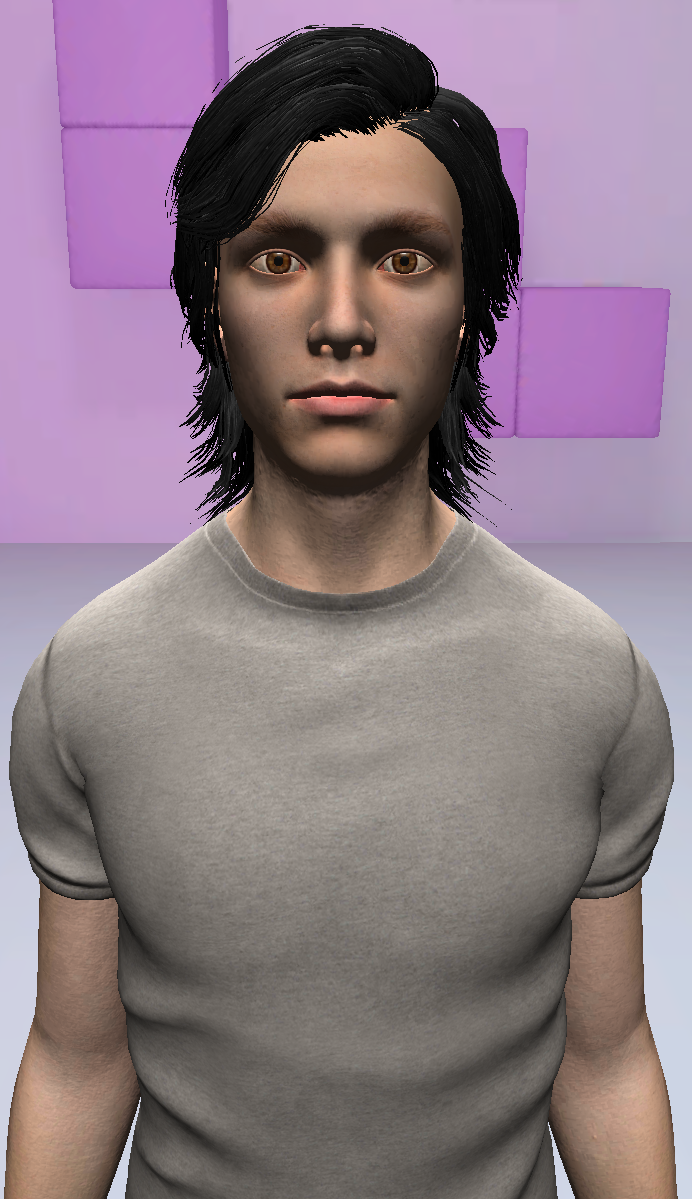}}
  \caption{}
  \label{fig:i1} 
\end{subfigure}
\addtocounter{subfigure}{2}
\hspace{-5pt}
\begin{subfigure}{.195\textwidth}
  \centering
  \frame{\includegraphics[width=\textwidth]{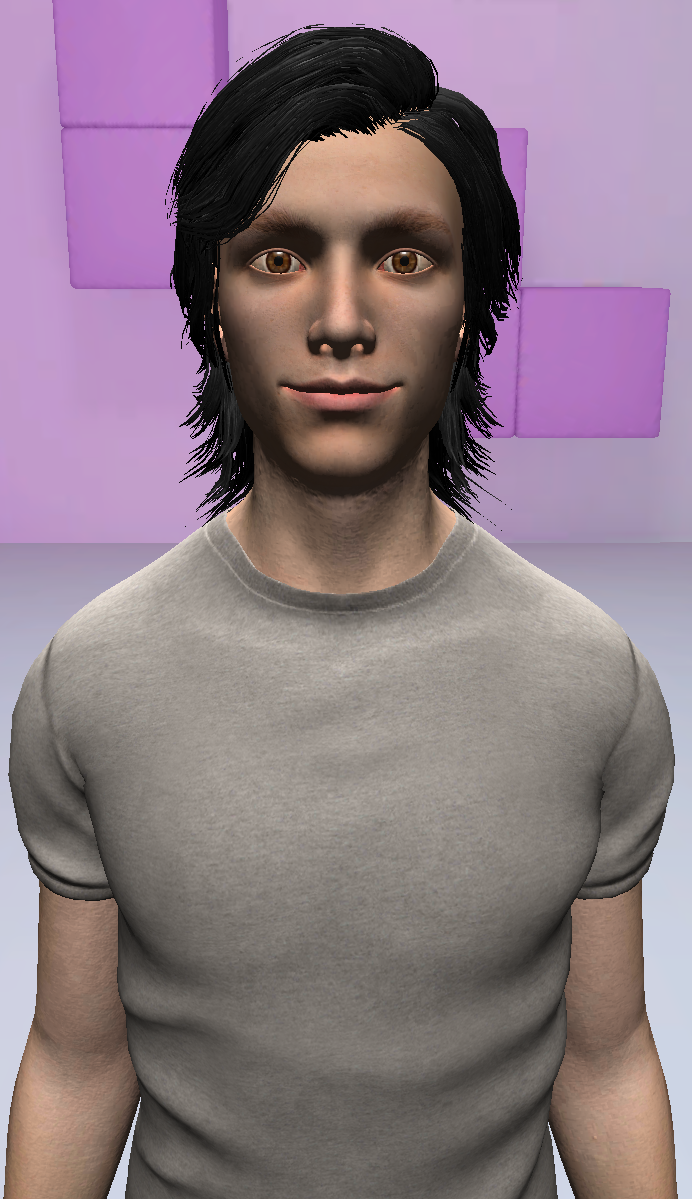}}
  \caption{}
  \label{fig:i4}
\end{subfigure}
\begin{subfigure}{.195\textwidth}
  \centering
  \frame{\includegraphics[width=\textwidth]{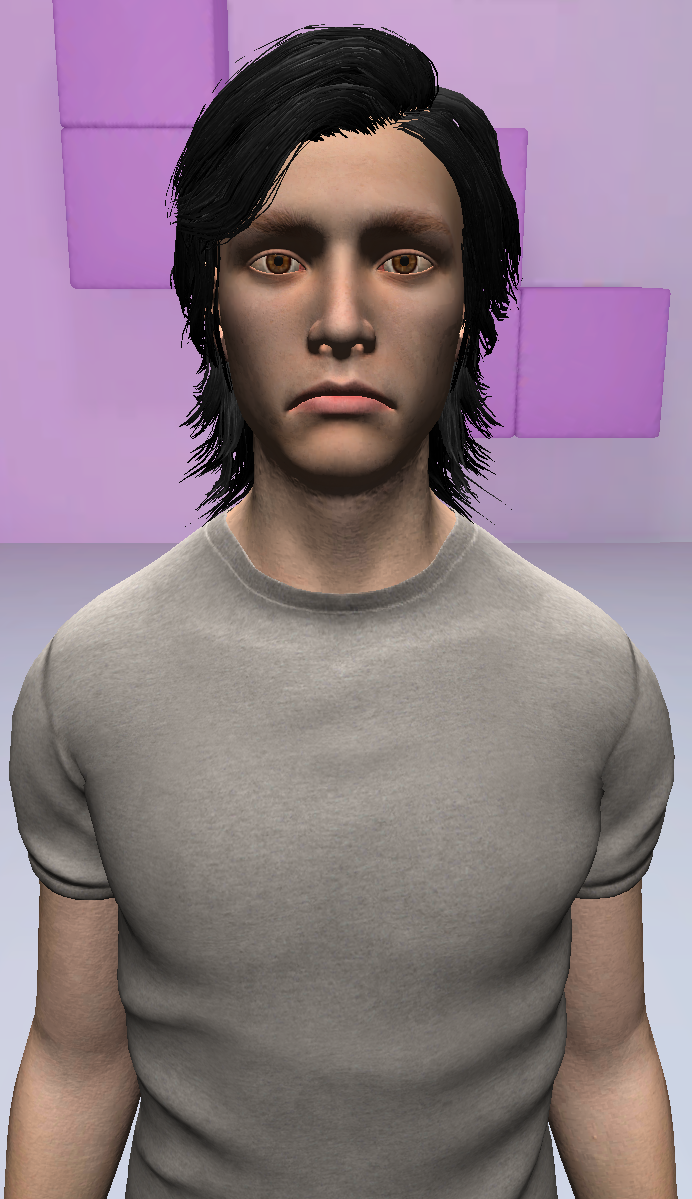}}
  \caption{}
  \label{fig:i5}
\end{subfigure}
\begin{subfigure}{.195\textwidth}
  \centering
  \frame{\includegraphics[width=\textwidth]{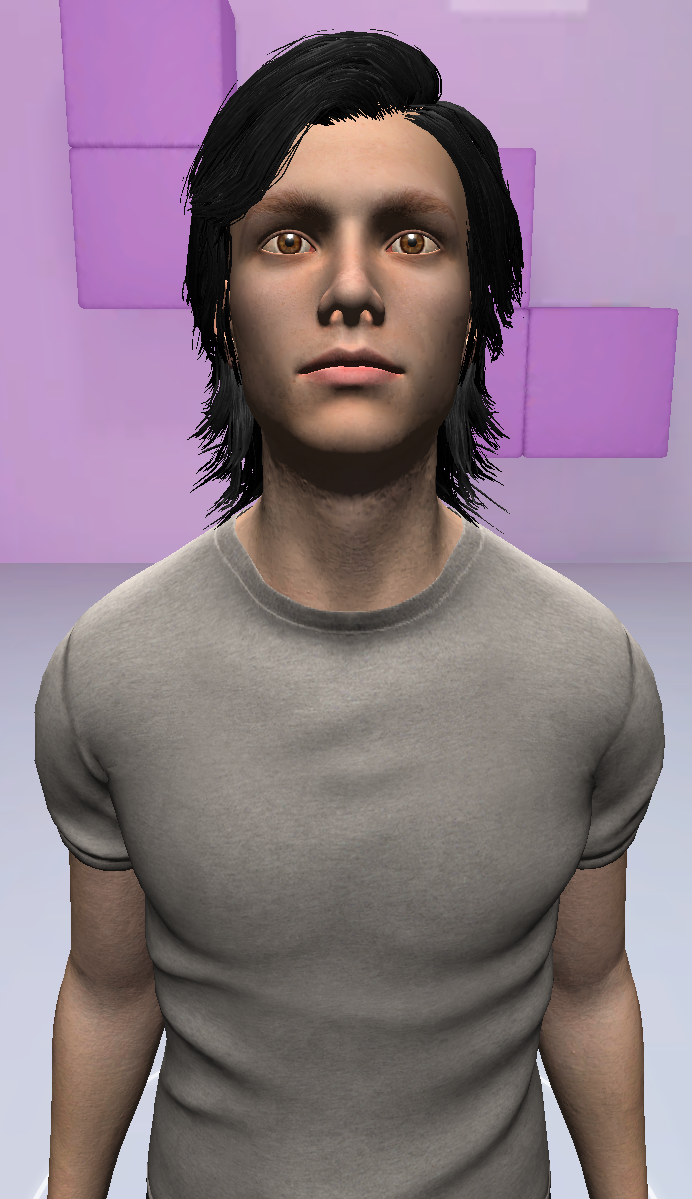}}
  \caption{}
  \label{fig:i6}
\end{subfigure}
\begin{subfigure}{.195\textwidth}
  \centering
  \frame{\includegraphics[width=\textwidth]{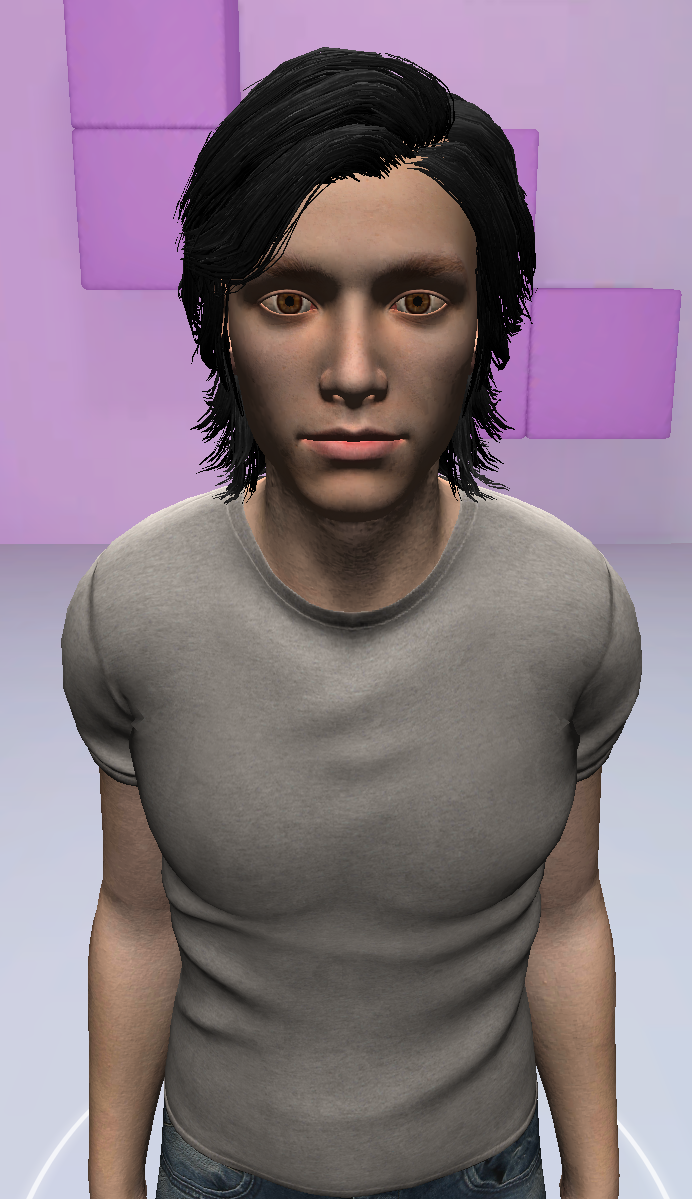}}
  \caption{}
  \label{fig:i7}
\end{subfigure}
\caption{Experimental manipulations for Study I, as described in Table~\ref{tab:independent}. Manipulations (b) and (c) used the same body as the control condition (a), with changes to speed of navigation.}
\label{fig:independent}
}
\end{figure*}
\section{Study I}

We conducted an outside-the-lab VR study, as research has proposed this to be a valid method for low-cost VR user studies with high power~\cite{steed2016,Mottelson:2017}.
Each participant was given a commodity cardboard VR headset to use in combination with his or her smartphone. This was deemed a cost-effective approach enabling a large-scale user study that would allow us to try many experimental conditions and, thereby, many avatar manipulations.

The overall purpose of this study was to i) validate the feasibility of influencing affect with avatar manipulations in VR, ii) experiment with several avatar manipulations with high power, and iii) identify specific questions pertaining to the body-ownership--affect link for further investigation via a laboratory study focusing on high internal validity.

With Study I, we were interested in core affect~\cite{ekkekakis2013measurement}. Since this is the broadest class of non-reflective affective feelings available to the consciousness, core affect is simpler to address than the underlying emotions and moods. Because it was not clear in advance how the manipulations would influence participants, we decided on a dimension-based conceptualization of core affect, involving valence and arousal (similar to Russell's circumplex model of affect~\cite{ekkekakis2013measurement}). 

\subsection{Participants}
In all, 207 undergraduate computer science students took part in the study. Most were male (51 females participated), and the age range was 19--50 ($M = 23,\ SD = 3.8$). Smartphones with iOS were slightly more common than Android ones among the participants, with 110 using iOS in the study. Participation was counted toward the students' credit for a compulsory introductory HCI course.

\subsection{Apparatus}
The participants used their smartphones within a Google cardboard VR headset that had a head strap. A few students did not have a compatible device, so we lent them one. The application was developed with the Google VR SDK for Unity 2017, deployed for Android and iOS and distributed via the relevant app stores. We employed an inverse kinematics (IK) system for the humanoid avatar, using the head as the only tracked point, with three degrees of freedom (DoF).

\subsection{Design}

We used three tasks in the design of our between-subjects experiment. The independent variable was body manipulation, with seven conditions (see Table~\ref{tab:independent} and Figure~\ref{fig:independent}): control, speed (\textit{slow}/\textit{fast}), face (\textit{smile}/\textit{frown}), and posture (\textit{upright}/\textit{hunched})

These specific avatar manipulations were chosen based on a review of body--affect relations. Coombes et al.~\cite{coombes2005}, for instance, show that speed of a task varies as a function of affect, and Wallbott~\cite{wallbott1998} presents an overview of how postures are related to emotion. Finally, research has showed how facial expressions can modulate affect~\cite{strack1988,niedenthal2007,Jun:2018}.

The tasks were adapted from pen-and-paper ones in Strack et al.'s work~\cite{strack1988} to suit an embodied VR experience. In all three tasks, the subject selected spheres in turn, on the basis of the number or letter printed on them. Each sphere was initially white, turned red when the subject looked at it, and then turned blue once the subject had been dwelling on it for two seconds. While participants were performing the tasks, a large mirror was in front of them, showing them their avatar (see Figure~\ref{fig:tasks}). Between tasks, the mirror served as a screen displaying buttons with which participants answered questions, again via two-second dwell times. The specifics of the tasks are described below.

\subsubsection{Task I}
Task I was designed to give familiarity with selecting targets by dwelling on them via head position and to introduce Likert-scale use in VR. Participants selected two spheres, numbered 1 and 2, by dwelling on each in turn. 

\subsubsection{Task II}
In the second task too, the participants selected numbered spheres, this time from 1 to 9. 

\subsubsection{Task III}
For the final task, they were asked to select all spheres with vowels (``Y'' was optional) on them, in any order. A sphere was present for each letter of the alphabet.

\begin{table}[h!]
\centering
\caption{The experimental manipulations in Study I (the hypotheses correspond to SAM responses).}
\begin{tabular}{llll}
&
\textbf{Variable} & 
\textbf{Manipulation} & 
\textbf{Hypothesis} \\ \hline
(a) & Control & None & Baseline \\
(b) & Rotation speed & Slow (-40\%) & High affect \\
(c) & Rotation speed & Fast (+40\%) & Low affect \\
(d) & Face & Smile & High affect \\
(e) & Face & Frown & Low affect \\
(f) & Posture & Upright & High affect \\
(g) & Posture & Hunched & Low affect \\
\end{tabular}
\label{tab:independent}
\end{table}
\noindent
In the control condition (a), there were no manipulations. For the speed conditions, we increased/reduced 40\% rotational gain around the $y$-axis, such that a 100-degree movement would result in rotation of 140 degrees for (b) and 60 degrees for (c). The face conditions manipulated facial features (using built-in morphing), to produce smiling (d) and frowning (e). Postural manipulations were implemented by adding artificial head and chest targets for the IK model, resulting in either upright (f) or hunched (g) posture.

\subsection{Measurements}
The primary dependent variables were valence and arousal, measured with a nine-point SAM (see Figure~\ref{fig:SAM})~\cite{SAM} administered with the textual prompt ``How do you feel?'' (Table~\ref{tab:dependent} provides the full list of dependent variables).
We chose the SAM to measure the primary dependent variables because it requires few of the cumbersome in-VR button selections, which would be prevalent using common affect questionnaires (e.g., PANAS~\cite{panas}). 
The dominance scale sometimes used as a third dimension for the SAM instrument is more emotion- than core-affect-oriented~\cite{ekkekakis2013measurement} so was not used in our study.  

\begin{table}[h!]
\centering
\caption{The dependent variables for Study I.}
\begin{tabular}{llll}
\textbf{Measurement} & \textbf{Instrument} & \textbf{Category} \\ \hline
Valence & SAM & Affect \\
Arousal & SAM & Affect \\
Difficulty & Likert scale & Post-task metric \\  
Completion time\ & Clock & Performance \\
Error rate & Head orientation\ & Performance \\
Mirror time & Head orientation & Activity \\
Body ownership & Two questions & Questionnaire \\
VR experience & Four options & Questionnaire
\end{tabular}
\label{tab:dependent}
\end{table}
\noindent

Also, we measured task difficulty (subjective difficulty), speed (task duration), and accuracy (error rate) for each task. In addition, we recorded the time spent looking in the virtual mirror, as the number of frames for which the participant's point of view collided with the mirror.  

\subsection{Procedure}
\begin{figure}[h!]
\begin{center}
\frame{\includegraphics[width=.7\textwidth]{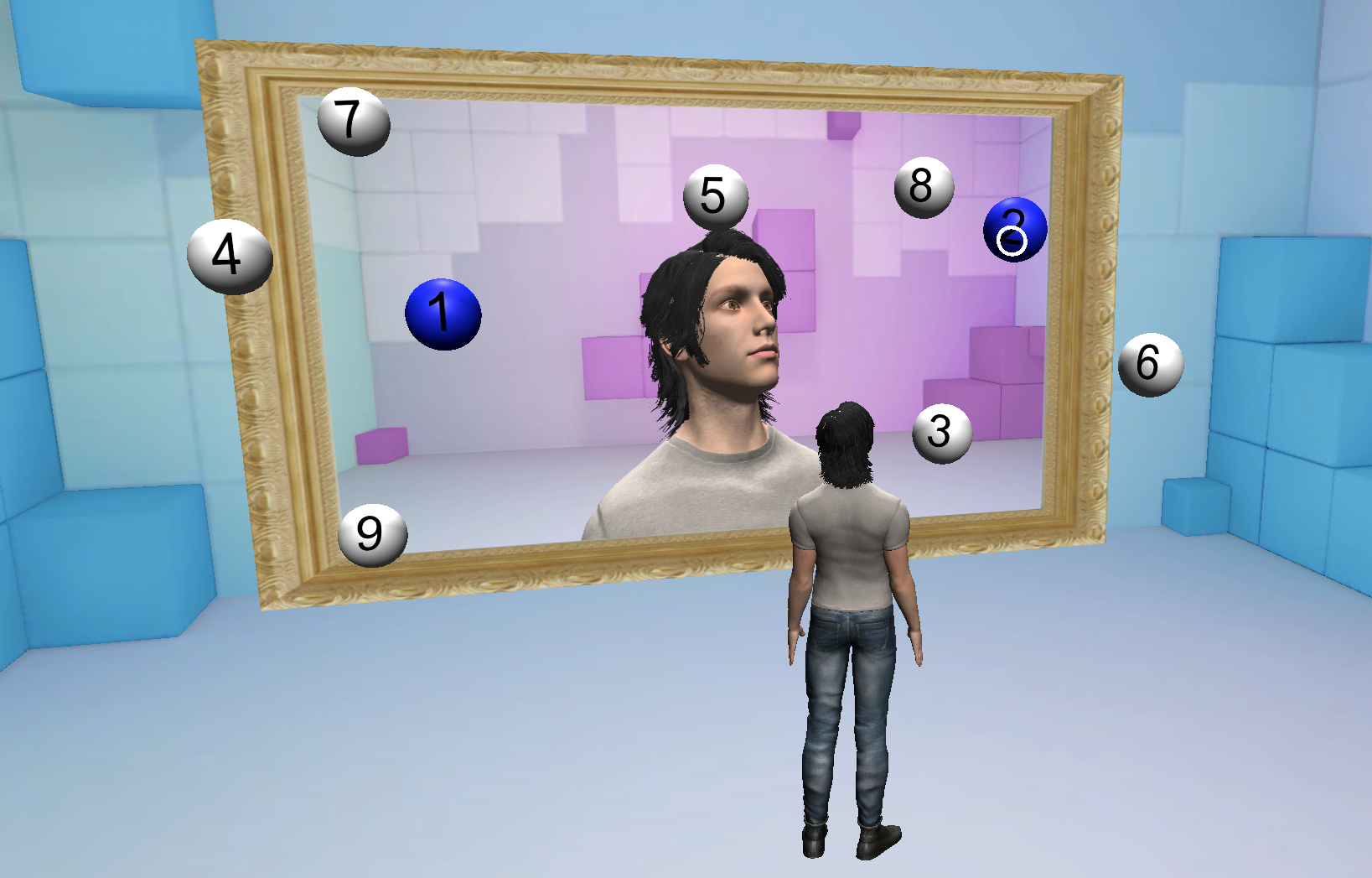}}
\caption{Task II shown from a third-person perspective -- the participant is placed in front of a mirror in a virtual room, and numbered spheres are selected in turn via dwelling. }
\label{fig:tasks}
\end{center}
\end{figure}

Participants entered a virtual room (see Figure~\ref{fig:tasks}) after installing and opening the mobile application and placing their phone in the cardboard VR headset. In this room, a standing sex-matched avatar could be seen in the mirror. Participants' coarse body movements were mapped in real time, with head rotation used alongside IK to create a sense of body ownership over the virtual character. Each participant was randomly assigned to one of the seven experimental conditions.
\noindent
Each subject performed, in all, three tasks, in a predefined order. In each task, the participants looked for floating spheres with predefined locations.
To create the illusion of body ownership, the mirror showed the avatar moving in visuo-motor synchrony with the participant.  

The participants rated the difficulty of each task upon its completion. After the third task, they filled out a SAM form that used a nine-point Likert scale created for VR.  


\subsubsection{Debriefing}  
After completing the VR part of the study, participants completed a brief Web-based questionnaire, providing data on body ownership and prior VR experience. We defined their body ownership as the maximum value reported for these two questions on body ownership used by Bakakou and colleagues~\cite{Banakou:2013}):
\begin{itemize}
\item How much did you feel that the virtual body you saw when you looked down at yourself was your own body?\vspace{1.5mm}
\item How much did you feel that the virtual body you saw when you looked at yourself in the mirror was your own body?
\end{itemize}
\noindent
Additionally, we asked the participants to guess the purpose of the study, for exclusion of anyone who suspected its true purpose (no one did). Each participant included the unique ID generated by the app for linking the app's study log with the post-experiment questionnaire. A week later, we revealed the purpose of the study to the student participants.  

\subsection{Results}

We tested the responses for normality, and a Shapiro--Wilk normality test showed that body-ownership, valence, and arousal responses did not follow a normal distribution. Therefore, our reporting of results in the following section refers to non-parametric statistics. Kruskal--Wallis tests were used, with $\chi^2$ test results reported in this section. For normally distributed data, $F$-scores from ANOVA tests are given.

\subsubsection{Data}

Most participants had only a little or some prior experience with VR; 21\% had never tried it before, 55\% had a few times, and 24\% had considerable VR experience.
We excluded two of the 207 participants for taking too long (>30 minutes), and 37 were removed from the sample for not experiencing body ownership (BO) with the virtual avatar (BO $\leq2$). Data from 168 participants remained.


The average time for completing the study was 297~s ($SD = 197$). No participants were excluded for guessing the purpose of the study; the vast majority responded with variations of ``I don't know,'' with some speculating that the aim was to understand usability aspects of VR navigation. A little less than half of the time, 119~sec. ($SD = 37$), was spent looking in the mirror. The mean difficulty (from 1 to 9) for each respective task was 1.8, 2.9, and 4.4, showing participants' ease with completing the tasks.

\subsubsection{Affect}

\begin{figure}[h!]
\begin{center}
\includegraphics[width=.5\textwidth]{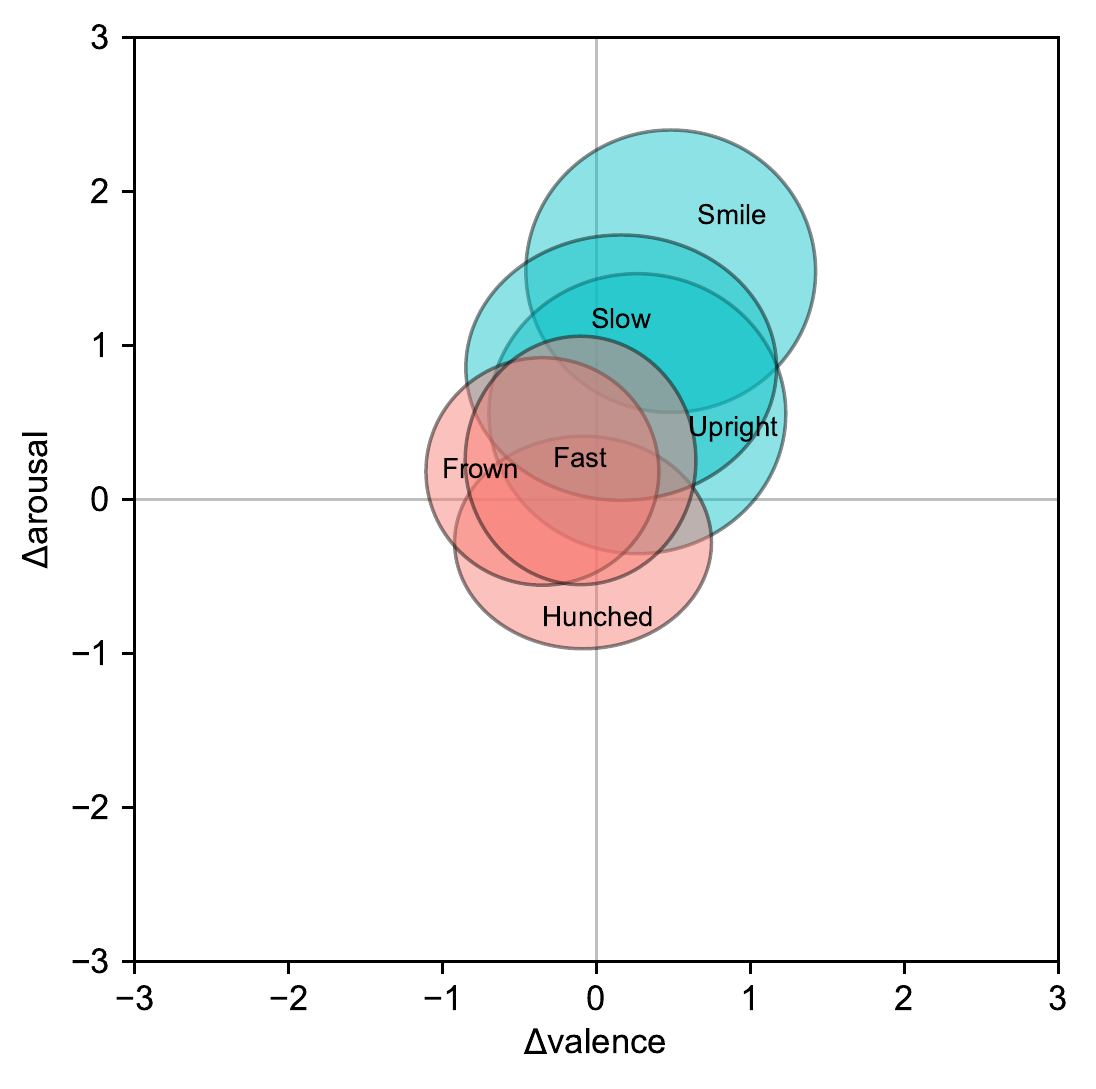}
\caption{Difference in means from the six conditions, normalized in terms of the control condition. Blue ellipses represent conditions with hypotheses of high affect, and red ones represent conditions with hypotheses of low affect. The ellipses are fitted to represent 95\% CIs.}
\label{fig:plot1}
\end{center}
\end{figure}

The control group (without avatar manipulations) reported, on average, a score of 5.3 for valence and 3.3 for arousal. Figure~\ref{fig:plot1} shows this group's differences in means for both valence and arousal; it is evident that the differences in valence between conditions are negligible (with means between 5.0 for frowning and 5.8 for smiling), while differences in arousal are more pronounced (means range from 3.0 for hunched posture to 4.8 for smiling).  

\noindent

We did find a significant effect of experimental condition on arousal: $\chi_6^2 = 14.61,\ p = .02$.
A Bonferroni-adjusted \emph{post hoc} Dunn's test showed that \textit{smile} and \textit{hunched} differed significantly, with $z = -3.32,\ p < 0.01$. We found no significant effect of experimental condition on valence: $\chi_6^2 = 2.75,\ p = .84$.  

The condition \textit{smile} showed itself to be the most effective manipulation for causing positive affect. Non-intuitively, its counterpart, \textit{frown}, did not emerge as the most effective contributor to negative affect; rather, it seems that this and other manipulations hypothesized to induce negative affect had little to no effect.

We noticed that positive avatar manipulations (\textit{smile}, \textit{slow}, and \textit{upright}) led to increased arousal relative to the control (see Figure~\ref{fig:plot1}). Comparing positive and negative manipulation reveals significance for arousal: $\chi^2_1 = 7.51,\ p < .01$.

Conversely, we observed negative manipulations (\textit{frown}, \textit{fast}, and \textit{hunched}) to have limited effect, with \textit{hunched} being the most effective. We noted that other researchers too have found positive affect easier to induce than negative affect; e.g., Schaefer et al.~\cite{schaefer2010} collected affect measurements for 64 movie clips and consistently found stronger induction of positive affect than creation of negative affect with these stimuli.

\subsubsection{Body ownership}

With regard to body ownership -- the degree to which sensory cues coalesce in the perception that a virtual body is ``my body''~\cite{kilteni2015} -- we expected to find consistent levels across all the manipulations. Therefore, we checked for an effect of experimental condition on level of body ownership. Indeed, we found no such significance: $\chi_6^2 = 6.61,\ p = .36$.

The results from Study I showed that body ownership varied significantly with the time spent looking in the mirror: $\chi_8^2 = 20.8,\ p = .008$. That is, the level of reported belief in the avatar being the participants' own body rose with the amount of time looking at the avatar in the mirror in front of them. This shows the importance of mirrors in body-ownership illusions, and it confirms that cardboard VR systems, modest fidelity notwithstanding, can induce those illusions for most participants (as suggested in earlier  work~\cite{Mottelson:2017,steed2016}).

\begin{figure}[h!]
\centering
\includegraphics[width=.35\textwidth]{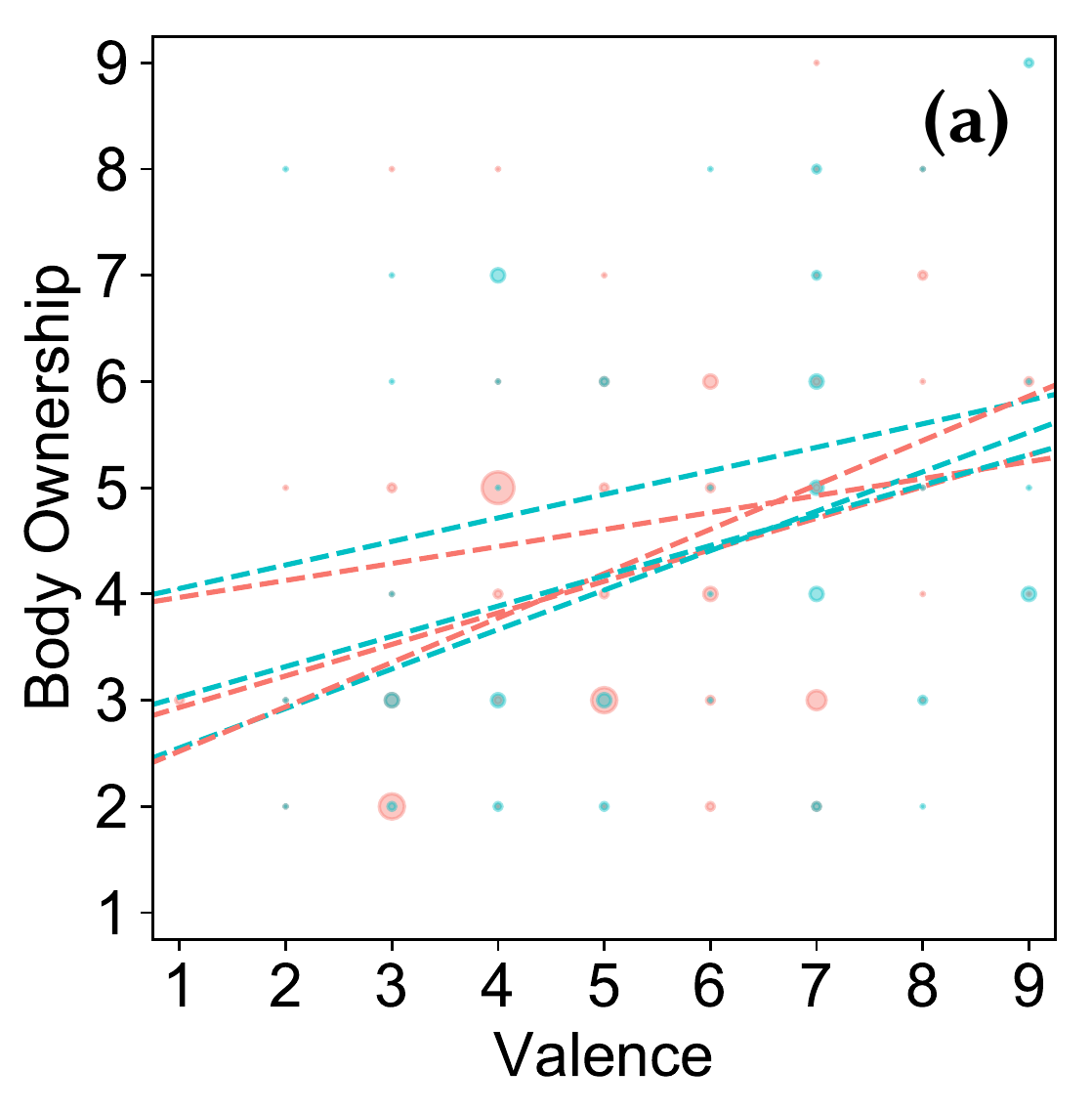}
\includegraphics[width=.35\textwidth]{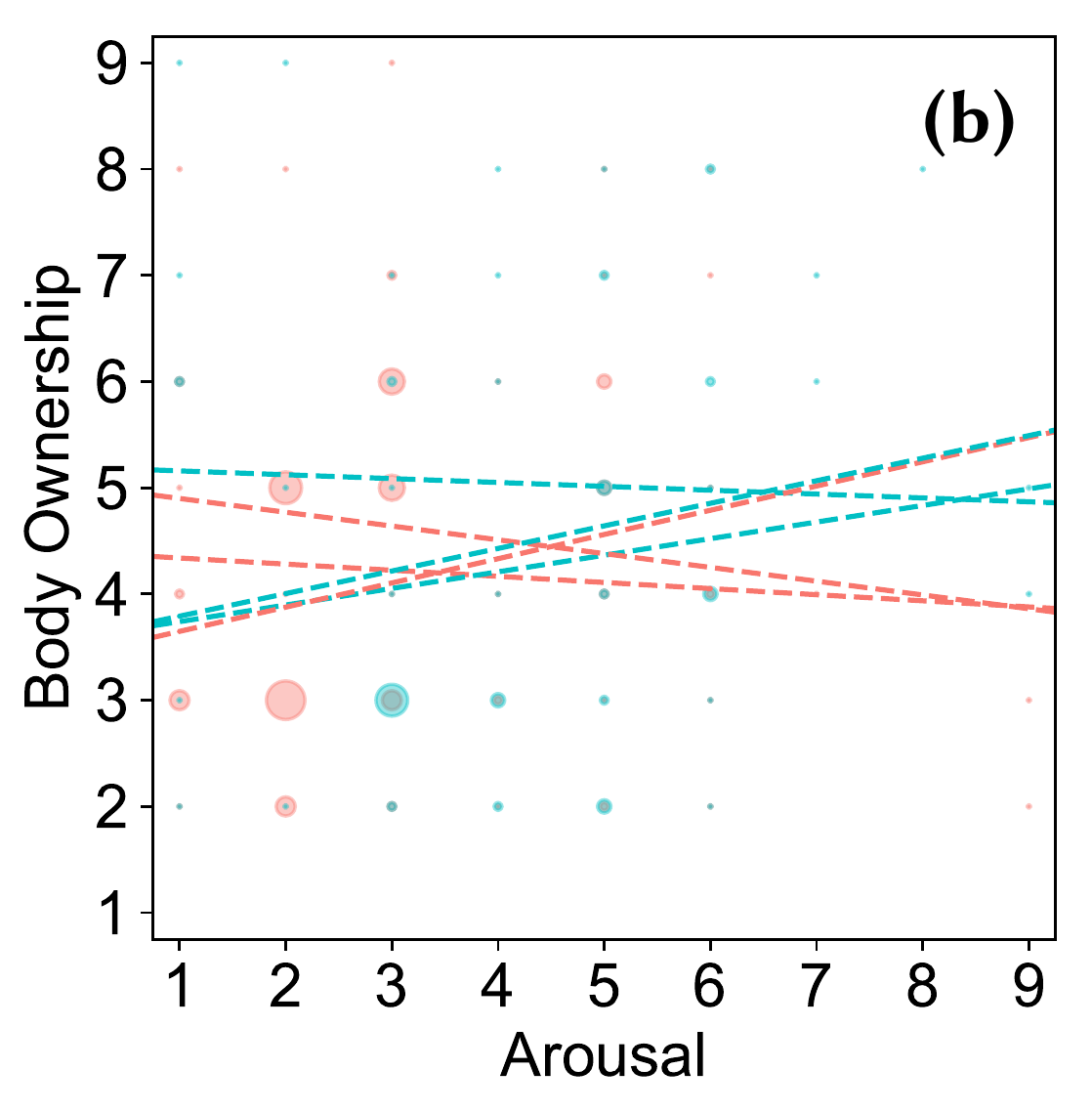}
\caption{Plots showing the relationship of body ownership with (a) valence and (b) arousal, where the size of the dots indicate frequency and the dashed lines represent the trend lines for positive (blue) and negative (red) manipulations.}\label{fig:spearman}
\end{figure}

A Kruskal--Wallis test showed that valence varied significantly with body ownership: $\chi_7^2 = 21.3,\ p < .01$. 
A Spearman's $\rho = .29$ showed that body ownership correlates somewhat with valence; 
the same was not true for body ownership and arousal, $\rho = .08$ (see Figure~\ref{fig:spearman}).

\noindent

\subsubsection{Performance}

Coombes et al.~\cite{coombes2005} described a link between task performance (speed/precision) and affect; specifically, they reported that lower affect leads to higher speed and error rate in a motor task. Therefore, we were interested in whether task performance would vary as a function of affect in this study.  

Participants completed the study most quickly in the \textit{frown} condition ($M=250$~sec.) and most slowly with \textit{smile} ($M = 353$~sec.); however, the speed difference attributable to avatar manipulation was not significant: $F(6, 161) = .61,\ p = .72$.

For error rate, we used the mean error rate over the three tasks, calculated thus:
for each task, we computed the Levenshtein distance, $lev(\alpha, \beta)$, between the optimal sequence of actions, $\alpha$ (e.g., Task II's $\alpha = $``$123456789$''), and the participant's sequence of actions, $\beta$.
Error rate was, similarly to speed, not found to vary significantly with avatar manipulation: $F(6,161) = 1.22,\ p = .30$.


\subsubsection{Summary}

The three avatar manipulations hypothesized to induce negative affect (\textit{fast}, \textit{frown}, and \textit{hunched}) did produce a lower average affect score than each of the manipulations hypothesized to induce positive affect (\textit{slow}, \textit{smile}, and \textit{upright}). This effect was statistically significant for arousal when the positive- and negative-condition groups were compared.
Also, an omnibus Kruskal--Wallis test showed significance for arousal, and a Bonferroni-corrected \emph{post hoc} test found the distributions to differ significantly for the pairing \textit{hunched} and \textit{smile}.

Body ownership significantly varied with valence; this was not the case for arousal.
We did not find evidence suggesting that body ownership is influenced by avatar manipulation. 

Neither did we find evidence for speed or error rate being influenced by the avatars' manipulation.

The results of Study I suggest that facial manipulations to avatars do alter affective responses, with posture manipulations having a similar but less pronounced effect. Manipulating speed does not seem to alter affective responses. Finally, our findings point to manipulations for positive affect as more efficient than those intended to induce negative affect.

\subsubsection{Discussion}

The mean score from the body-ownership reports (scale: 1--9) was 4.36 ($SD = 1.91$). While this is not unusually low, it does suggest that something in the circumstances of Study I limited the body-ownership scores. We believe the factors might include i) the commodity VR equipment with only 3 DoF; ii) the short study duration (five minutes, with two minutes of mirror time); and iii) the lack of internal control in outside-the-lab experimentation.

While outside-the-lab experimentation allows for rapid implementation of large-scale user studies at low cost, it imposes design constraints and creates practical limitations to the experiment. In particular, lengthy studies work poorly outside a laboratory setting, and adherence to the protocol (e.g., standing up and using a head strap) is hard to confirm. Additionally, it is rendered difficult to obtain reliable measurements of bodily aspects, such as posture and locomotion.
We addressed these concerns by conducting a lab-based study, with higher internal validity and technical fidelity.

\section{Study II}

The purpose of Study II was to address unresolved questions from Study I about i) the interplay between body ownership and affect and ii) the effect of posture changes on affect.

Since valence varied with body ownership in Study I, we wanted Study II to cast more light on the connection between body ownership and affect, through measurements with high internal validity. While Study I showed only a non-significant difference between the \textit{hunched} and \textit{upright} conditions in terms of affect, that study was limited in a number of respects. With Study II, we hoped to ascertain whether a more elaborate setup and rigorous study protocol would reveal differences in affective responses between posture changes to the avatar.


\subsection{Participants}
We had 42 participants in this study, with an age range of 21--34 ($M = 26.3,\ SD = 3.4$). We recruited people to take part via an internal mailing list. The first participant was excluded on account of technical errors, so we report on analysis performed for 41 people (22 of whom were female). All participants were given a gift worth the equivalent of \$20~USD for their time. Participants signed a consent form before the experiment commenced. No one who took part in Study I took part in Study II.

\subsection{Apparatus}

\begin{figure}[b!]
\centering
  \frame{\includegraphics[width=.7\textwidth]{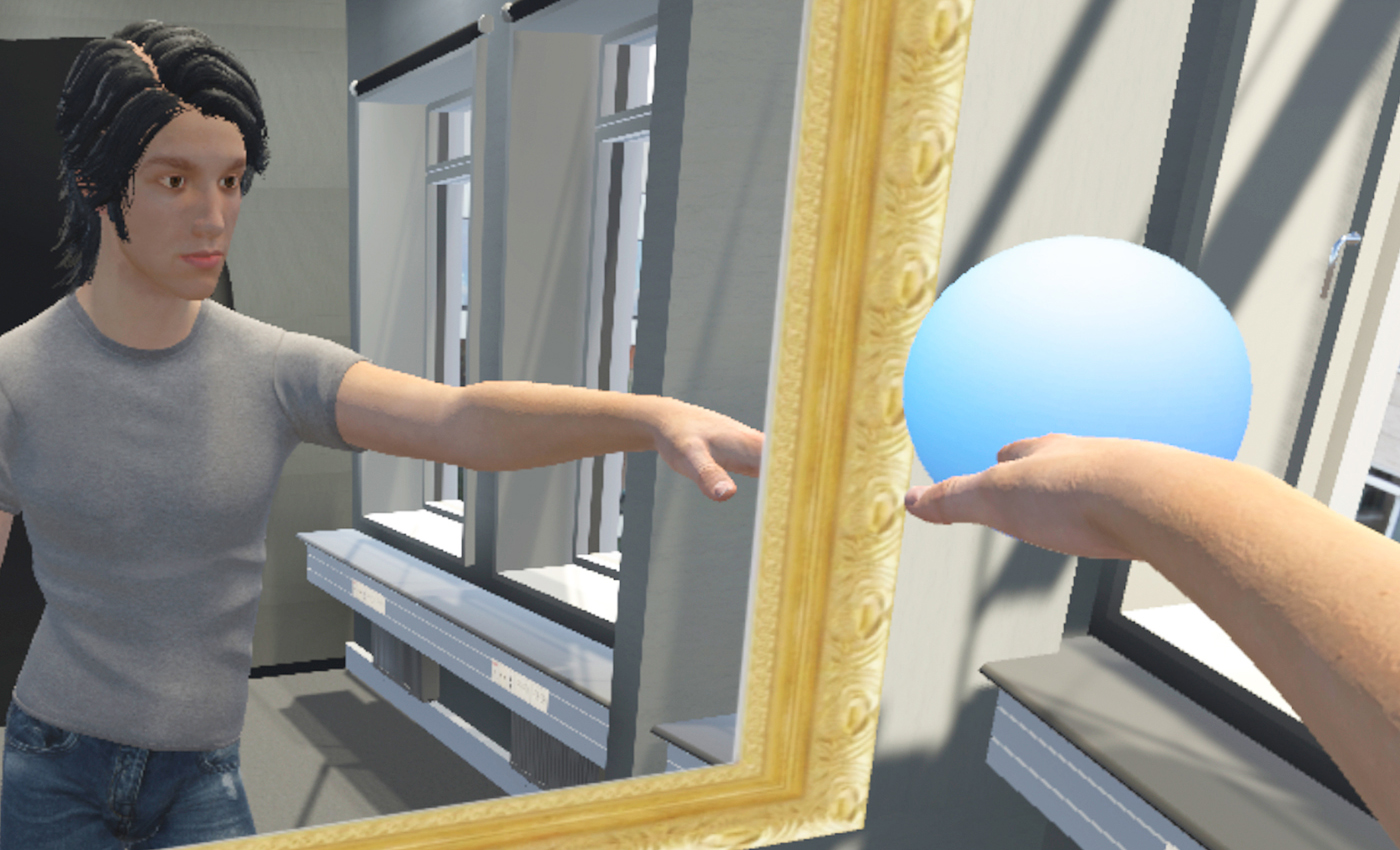}}
\caption{The view of the participant immersed in the VR world.}\label{fig:studyii}
\vspace{-1.5mm}
\end{figure}

For this study, we used an HTC VIVE system (6 DoF) in combination with an OptiTrack motion-capture system for state-of-the art body-tracking. 
We employed a Unity scene similar to that developed for Study I, using a desktop PC (2.8~GHz Intel i7, 12~GB RAM, NVIDIA GTX 980), running Windows 10 Pro.
Tracking was performed with Motive, using eight Prime~13 cameras at 120~Hz (the same frame rate used for HTC VIVE lighthouses), positioned in a semicircle (see Figure~\ref{fig:studyii}). We tracked the hands, elbows, feet, chest, and shoulders. Retroreflective markers were attached to the head-mounted display for SteamVR--Motive alignment. The chest and shoulders were not attached to the IK system but were tracked for later analysis.  

\subsection{Design}

Study II employed a between-subjects design with avatar manipulation as the independent variable (there were three groups, with 14 people in each). Because the manipulations of avatar posture in Study I exhibited no clear effect, we were interested in seeing whether this was due to the lack of technical fidelity (e.g., DoF, latency, and screen-resolution issues) and experimental control (e.g., issues with supervision, whether subject were standing up, adherence to procedure, and interruptions). On this basis, these three conditions were chosen for the independent variable: \textit{hunched}, a control, and \textit{upright}. The last was added halfway through the study without the experimenter's knowledge; this was done since the analysis in Study I showed little difference between the control and the \textit{hunched}-condition group.

We chose PANAS results for the dependent variable of affect -- this measure has consistently been shown to have high validity~\cite{panas2}. The instrument offers a broader conceptualization of affect than in Study I, in that it features items related to emotion and mood, not only core affect~\cite{ekkekakis2013measurement}. Thereby we hoped to gain a more nuanced picture of the influence of our manipulations while retaining a dimensional view of affect. The PANAS instrument was administered on a computer alongside the textual prompt ``Please indicate to what extent you feel this way right now''.

We chose a task similar to those in Study I, although this one was longer and required full-body movements instead of only head orientation and dwelling. Again, a mirror was present, in which the visuo-motor synchronous, sex-matched avatar was visible. The virtual room where participants were immersed was a replica of the physical room in which the experiment took place (see Figure~\ref{fig:studyii}). The experiment was conducted by someone aware of neither the study's purpose nor of Study I.\\[.25cm]
In summary, in comparison to Study I, Study II had
\begin{itemize}
\item higher VR fidelity (HTC VIVE instead of cardboard),
\item full-body tracking (8 $\times$ OptiTrack Prime 13),
\item longer duration (17 min., as opposed to 5 min.),
\item an extensive construct for affect (PANAS, not SAM),
\item fewer conditions (three instead of seven), and
\item fewer participants (42 instead of 207).
\end{itemize}
\vspace{-1mm}

\subsection{Procedure}

After calibration in which participants' bodies were aligned with their virtual avatar, the study proper began.
The study progressed with a series of floating 3D objects (spheres, cubes, and icosahedra) that disappeared when the participant tapped them with either hand. The experiment ended once the participant had tapped 200 objects. All objects were spawned in random locations between the participant and the mirror, such that the participant faced in the same direction with respect to the mirror throughout the study (see Figure~\ref{fig:studyii}). Thus, each subject was required to glance around, move about, and tap objects close to both the floor and the ceiling. This movement was reflected in the mirror placed in front of the participant.
After finishing the VR task, participants filled out a computer-administered post-experiment questionnaire in which i) PANAS results, ii) body-ownership data, and iii) demographic details were collected.




\subsection{Results}

\subsubsection{Data}

Participants used, on average, 17.6 minutes ($SD = 2.4$) for the study, with 14 minutes spent looking in the mirror ($SD = 1.7$).
The median score from the body-ownership reports (scale: 1--9) was 6 ($IQR = 4, 7$).
Most subjects had little prior experience with VR: 38\% had never tried it before, 38\% had tried it a few times, and 24\% had considerable VR experience.

\subsubsection{Affect}
\begin{figure}[h!]
\begin{center}
\includegraphics[width=.5\textwidth]{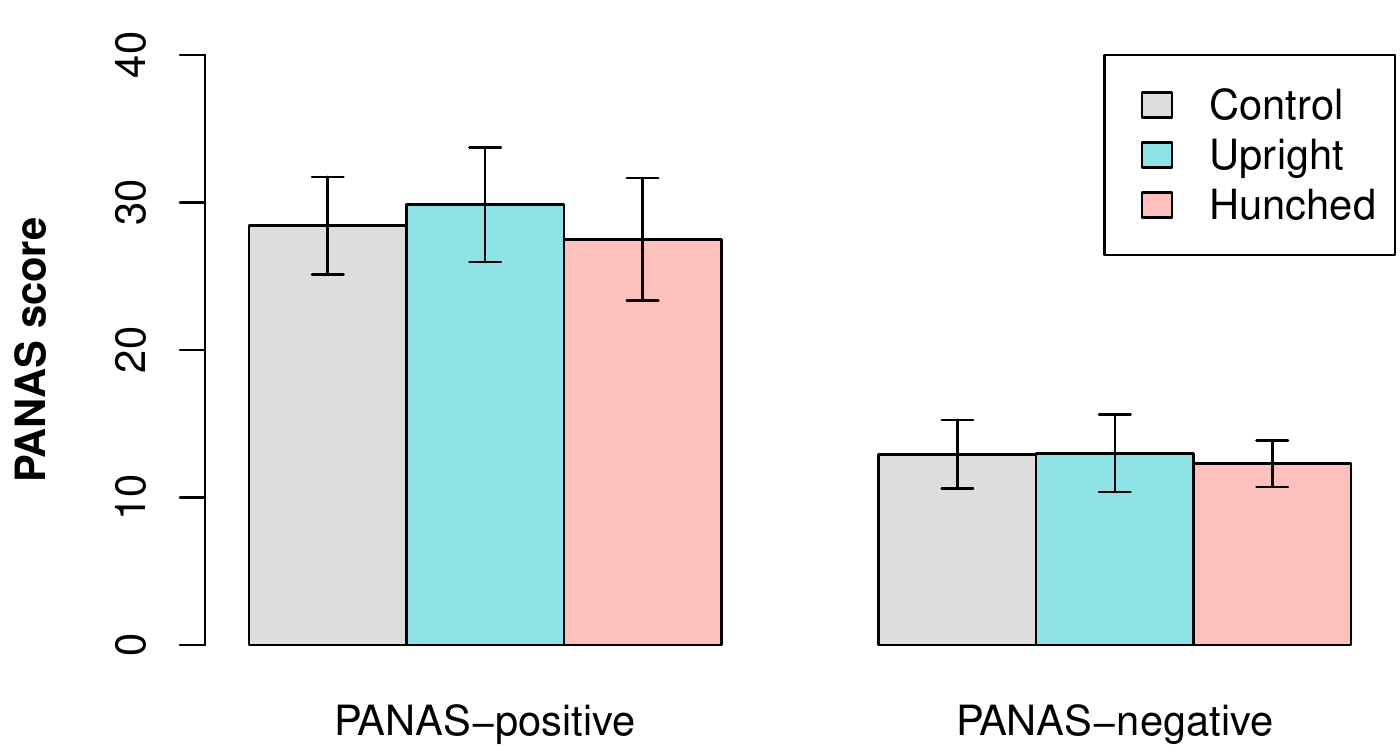}
\caption{Mean PANAS scores. Error bars show 95\% CIs.}  
\label{fig:study2bar}
\end{center}
\vspace{-1.5mm}
\end{figure}

The PANAS instrument covers two components, positive and negative affect, generating a score between 10 and 50 for each. These are considered two independent measures of affect. Figure~\ref{fig:study2bar} shows the negligible difference in affective responses between conditions.
We were unable to find any significant effects of avatar manipulations on either PANAS component: for the positive one, $\chi^2_{2} = .25,\ p = .88$; for the negative one, $\chi^2_{2} = .09,\ p = .95$.


\noindent


\subsubsection{Body ownership}

We did not find an effect of avatar manipulation on body ownership: $\chi^2_2 = 2.93,\ p = .23$.

Since valence was found to vary with body ownership in Study I, we hoped to gain a better sense of the body-ownership--affect relationship with this study. We tested whether the PANAS components varied with body ownership and found a significant effect of body ownership on the positive one: $\chi^2_7 = 14.92,\ p = .04$. Significance was not found for the negative component: $\chi^2_7 = 8.33,\ p = .3$.

\begin{figure}[h!]
\begin{center}
\includegraphics[width=.7\textwidth]{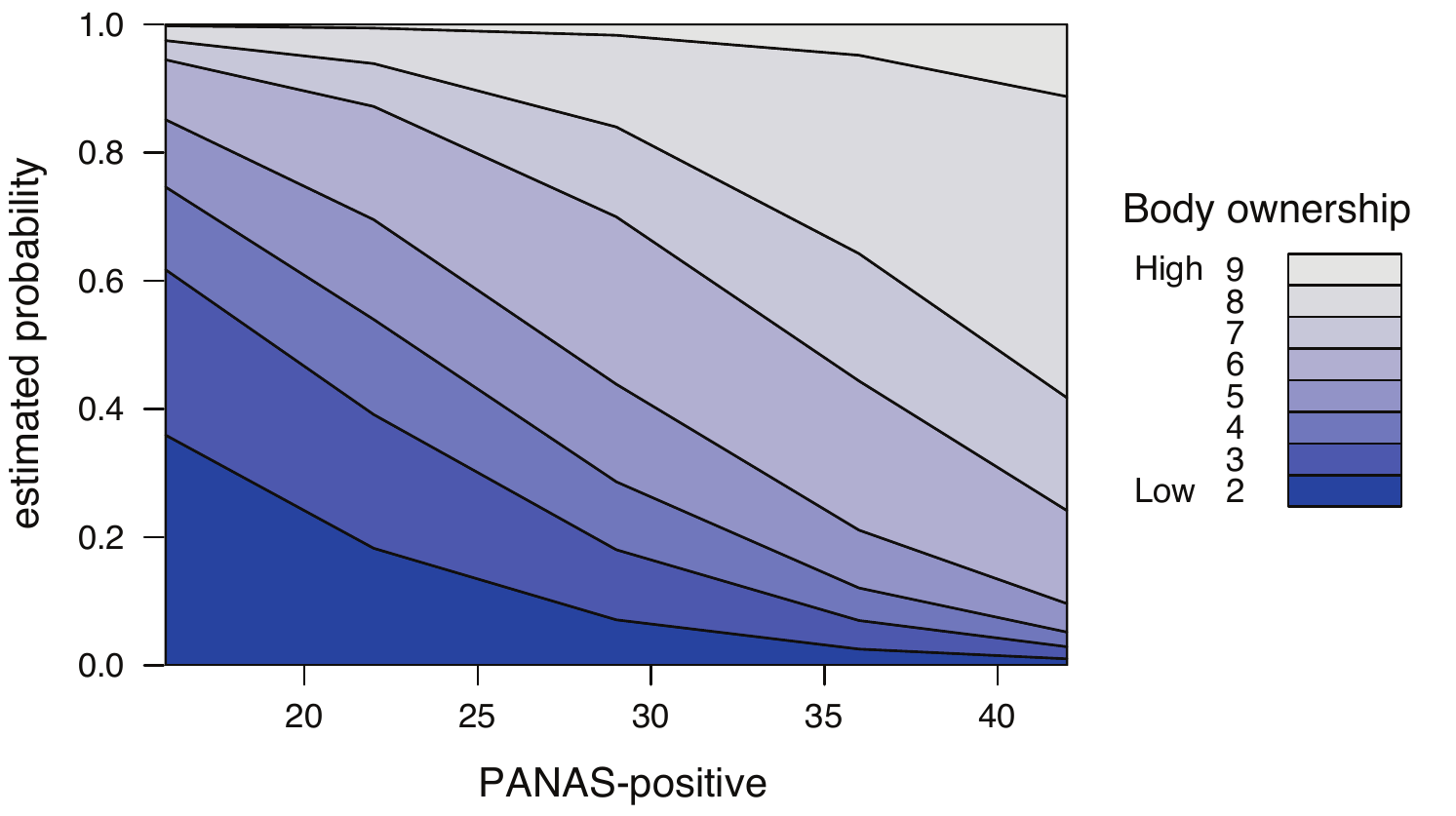}
\caption{The probabilities estimated from the fitted values of an ordinal logistic regression of body ownership on the positive PANAS-response component.}  
\label{fig:logreg}
\end{center}
\end{figure}

Inspired by Kilteni et al.~\cite{kilteni2015}, we treated body ownership as an ordered categorical value and in a combination with the numerical PANAS positive component we performed an ordinal logistic regression of body ownership. This yielded a fit value with a positive coefficient (the higher the positive-component response, the greater the likelihood of a high level of body ownership at the time). Figure~\ref{fig:logreg} shows the estimated probabilities from the logistic fit $P($\emph{Body ownership}$|$\emph{Pos. PANAS}$)$. The probability of high ownership increases as PANAS values rise.  
For instance, the estimated probability of the ownership score being $\geq7$ is $.76$ for a PANAS value of $42$ but only $.05$ for a PANAS figure of $16$. Ownership scores below 3 are most likely to be seen with PANAS scores under $16$, with an estimated probability of $.62$.

\noindent

Study I found valence to vary significantly with body ownership, while the foregoing analysis shows that the probability of high ownership increases with higher positive-affect scores. 
Together, these findings provide evidence that body ownership is an important factor in controlling the valence of affect in embodied VR experiences, particularly for inducing positive affect. 

We did not find a link between negative affect and body ownership, and 
we speculate that the low variance between subjects for the negative PANAS component reflects this: negative-component scores varied within the range 10--26 ($SD = 3.7$), while positive-component ones ranged from 16 to 42 ($SD = 6.4$).  

\subsection{Summary}

Study II did not reveal differences in affective responses between avatar conditions (control, \textit{upright}, and \textit{hunched}), just as Study I revealed no differences in affect attributable to full-body avatar manipulations. Hence, it is likely that posture changes do not have any influence on affect.

We found that body ownership varies significantly with positive affect. With higher affect scores, there was greater likelihood of high ownership values being reported: a positive-component PANAS score of $42$ (out of $50$) yields an estimated $.76$ probability of an ownership score of at least 7 (out of 9); a positive-component score of $16$ yields an estimated $.05$ probability of an ownership score of 7 or above.  


\section{Discussion}




Ba\~{n}os et al.~\cite{banos2004} found that affective content has an impact on presence in virtual environments: they reported that in non-affective environments presence depended mainly on immersion, while the relationship proved more complex for environments with affective content. Jun et al.~\cite{Jun:2018} too reported on presence and affect: with a body-ownership illusion they found presence to be positively correlated with valence. Specifically, the authors showed that owning an avatar with a happy face leads to higher presence estimates.

The study reported upon here expands our understanding of this interplay between virtual selves and perception. Firstly, we found that facial features are, in fact, efficient at influencing affect, while a weaker effect of this sort was found for upper-torso manipulations and movement speed. Secondly, we showed that positive affect is a good predictor of the likelihood of high body ownership. 
Our results suggest that valence is positively correlated with body ownership, while arousal is not. The results reveal, in addition, that positive affect is an important factor in body ownership. These findings are important for research considering affect and VR. For instance, mood-induction procedures in VR are likely to interfere with body ownership; study designs should be sensitive to this issue. 


Our examination of the link between virtual bodies and users' affect was driven by a desire to inform design. We believe that the studies speak to this goal well -- they may fruitfully inform the design of avatars and aid in creating a more solid foundation for subtle mood-induction techniques for virtual reality. For the domain of avatar design, our work suggests that some manipulations may influence affect, though not necessarily in great magnitude. As hinted at in prior work on VR~\cite{Jun:2018}, the most promising route for influencing affect seems to lie in manipulating facial expressions, especially smiling. It is far from clear that such manipulations work well for negative affect, however. Furthermore, indirect manipulations of affect seem more difficult to achieve; movement speed and posture produce unclear results. Further utility of the findings can be found in mood induction, as noted above. Existing mood-induction procedures for VR change the entire environment~\cite{banos2006, felnhofer2015, riva2007}. Our results show that less blatant manipulations might be feasible, although the interplay with body ownership suggests that their design might be difficult.

Study II found only the positive PANAS component, not the negative one, to vary with body ownership. While this may seem odd or counter-intuitive, the two components should be considered wholly independent. In a similar vein, Study I's manipulations intended to induce positive affect (\textit{smile}, \textit{slow}, and \textit{upright}) were more effective than their negative counterparts (\textit{frown}, \textit{fast}, and \textit{hunched}).
Also, in Study I only valence (not arousal) was found to respond to the condition. Both studies suggest that avatar manipulations are effective primarily for influencing positive affect.

\section{Conclusion}

We examined whether it is possible to influence affective responses by altering avatars in virtual reality. Results from an outside-the-lab study with 207 participants showed that this indeed is feasible. Manipulations to the avatars' facial features proved effective in modulating valence responses.

Moreover, we tackled the seemingly harder question of how affect interacts with the illusion of owning a virtual body. Our results show that positive affect is of great importance for body ownership: in Study I, valence varied with body ownership, with positive affect being found to follow body ownership. Our analyses show that high positive-affect responses increase the probability of high body-ownership responses. Together, these findings contribute substantially to our understanding of how emotion information influences fundamental VR constructs.  
\clearpage

\bibliographystyle{unsrt}  
\bibliography{references}  

\end{document}